\documentclass[onecollarge]{svjour2}

\usepackage{latexsym}
\usepackage{graphicx}
\usepackage{amsbsy,amssymb,amsmath}
\usepackage{mathptmx}

\journalname{Acta Mechanica}

\begin{document}

\title{Magnetic spherical Couette flow in linear combinations of axial
       and dipolar fields}
\author{Xing Wei \and Rainer Hollerbach}
\institute{Xing Wei \at Institute of Geophysics, ETH, Zurich, 8092,
           Switzerland \\ \email{xing.wei@erdw.ethz.ch}
      \and
           Rainer Hollerbach \at Department of Applied Mathematics,
           University of Leeds, Leeds, LS2 9JT, United Kingdom \\
          \email{rh@maths.leeds.ac.uk}}
\date{Received: date / Accepted: date}
\maketitle

\begin{abstract}
We present axisymmetric numerical calculations of the fluid flow
induced in a spherical shell with inner sphere rotating and outer
sphere stationary.  A magnetic field is also imposed, consisting of
particular linear combinations of axial and dipolar fields, chosen to
make $B_r=0$ at either the outer sphere, or the inner, or in between.
This leads to the formation of Shercliff shear layers at these particular
locations.  We then consider the effect of increasingly large inertial
effects, and show that an outer Shercliff layer is eventually
de-stabilized, an inner Shercliff layer appears to remain stable, and
an in-between Shercliff layer is almost completely disrupted even
before the onset of time-dependence, which does eventually occur
though.
\end{abstract}

\section{Introduction}
Spherical Couette flow is the flow induced in a spherical shell by
differentially rotating the inner and/or outer spheres.  Magnetic
spherical Couette flow, first considered by \cite{Holl94}, is the
magnetohydrodynamic analog, in which the fluid is taken to be
electrically conducting, and a magnetic field is externally imposed.
One motivation for studying this problem is to understand the dynamics
of the Earth's molten iron outer core, and the magnetic field that is
generated within it.  There have also been two experiments on magnetic
spherical Couette flow, the first in a uniform axial field imposed via
Helmholtz coils from the outside \cite{Dan}, and the second in a dipolar
field imposed via a permanent magnet embedded in the inner sphere
\cite{DTS1,DTS2}.

The question of what type of field to impose is indeed fundamental in
determining much of the subsequent dynamics.  Even if we insist that this
field be axisymmetric about the rotation axis of the inner and/or outer
spheres (otherwise the problem immediately becomes fully three-dimensional),
one could still impose an axial field from outside, or a dipole (or
quadrupole) from inside, or indeed any combination of these.  And because
of the highly anisotropic nature of the magnetic tension force, coupling
the fluid along the magnetic field lines, many of the results can be
explained largely by the geometry of the imposed field.  Fluid on field
lines that thread only the inner or only the outer sphere will co-rotate
with that sphere, whereas fluid on field lines linking the two spheres will
rotate at some intermediate rate.  The result is a shear layer, a so-called
parallel or Shercliff layer, on the particular field line separating one
type from another \cite{Star97,Star98}.  The thickness of these layers
scales at $Ha^{-1/2}$, where the Hartmann number $Ha$ is a measure of the
strength of the imposed field.

Another issue that turns out to be surprisingly important concerns the
nature of the inner and outer boundaries, and whether they are taken to
be insulating or conducting.  One obtains a standard Shercliff layer
only if both spheres are insulating.  For a conducting inner sphere,
\cite{Dormy1,Dormy2,Mizerski} showed that instead of merely a shear layer,
one obtains a super-rotating region, in which fluid rotates slightly faster
than either the inner or outer spheres.  And finally, if both boundaries
are conducting, this super-rotation can be made arbitrarily large, scaling
as $O(Ha^{1/2})$ instead of $O(1)$ \cite{Holl00a,Buehler}.  These two types
of super-rotation were both obtained for dipolar fields.  For axial fields,
one obtains a counter-rotation, which is again either $O(1)$ if only one
boundary is conducting, or $O(Ha^{1/2})$ if both are \cite{Holl00a,HS01}.

Motivated by these surprisingly different results for axial and dipolar
fields, \cite{Holl01} considered linear combinations of the two, and
showed that one may have super- and counter-rotating regions simultaneously.
Furthermore, because axial and dipolar fields both have the same $\theta$
structure, differing only in $r$, one can adjust the particular linear
combination so that the radial component of the field is zero on any
desired radial shell.  One can therefore choose the combination such that
$B_r=0$ on the outer sphere, or on the inner, or any radius in between.
One can thereby precisely control the geometry of the imposed field:  If
$B_r=0$ on the outer sphere, then all field lines thread only the inner
sphere, and vice versa.  If $B_r=0$ at some intermediate radius, then some
field lines thread only the inner sphere, and some only the outer.  In no
case though will there be any field lines linking the inner and outer
spheres.  In terms of the results discussed above, these particular linear
combinations are therefore important special cases.

Having $B_r=0$ on either the inner or outer boundary is also an important
special case in terms of the structure of the Hartmann boundary layers that
are generally expected to form there.  It is known \cite{Roberts,Loper}
that the standard Hartmann layer breaks down at isolated points where
$B_r=0$, so if $B_r=0$ everywhere on a particular boundary, the entire
structure of the boundary layer there will be altered, becoming essentially
a Shercliff layer that just happens to coincide with a boundary.  It is
for both of these reasons that these particular linear combinations of
fields are worthy of special attention.

All of these special geometrical properties discussed so far are in the
essentially linear regime, where the inertial term $\bf U\cdot\nabla U$ is
negligible, because the differential rotation between the two spheres is
infinitesimal.  It is obviously of interest then to consider what happens
when this term is no longer negligible.  At this point axial and dipolar
fields again differ quite dramatically.  For axial fields, the axisymmetric
basic state is largely unaffected, and one is interested instead in the
onset of non-axisymmetric instabilities \cite{HS01,Xing,Holl09}.  In contrast,
for dipolar fields, the basic state itself is strongly altered, with these
particular field lines distinguishing one region from another becoming
much less prominent \cite{HCF07}.

In this work then we will essentially combine \cite{Holl01} and \cite{HCF07},
taking the imposed fields to be these special linear combinations considered
by \cite{Holl01}, but extending this into the strongly nonlinear regime as
in \cite{HCF07}.  We will find the same general types of boundary layer
instabilities as in \cite{HCF07}, but will also show that having $B_r=0$ at
one boundary or the other, or in between, can still have a significant
influence on the solutions.

\section{Equations}

The equations to be solved are essentially the same as in \cite{HS01,Holl09},
$$\frac{\partial{\bf U}}{\partial t}+Re{\bf U\cdot\nabla U}=-\nabla p
  +\nabla^2{\bf U}+Ha^2(\nabla\times{\bf b})\times{\bf B}_0,\eqno(1)$$
$${\bf 0}=\nabla^2{\bf b}+\nabla\times({\bf U\times B}_0),\eqno(2)$$
except that ${\bf B}_0$ will no longer be a purely axial field.

Length is scaled by the radius of the inner sphere $r_i$, time by the viscous
diffusive timescale $r_i^2/\nu$, where $\nu$ is the viscosity.  $\bf U$ is
scaled by $\Omega r_i$, where $\Omega$ is the imposed rotation rate of the
inner sphere (with a stationary outer sphere).  The Reynolds number $Re=
\Omega r_i^2/\nu$ is then the non-dimensional measure of $\Omega$.  The
Hartmann number $Ha=B_0 r_i/\sqrt{\mu\rho\nu\eta}$ similarly measures the
strength of the imposed field $B_0$, where $\mu$, $\rho$, and $\eta$ are
respectively the permeability, density, and magnetic diffusivity.

These equations have been formulated in the limit of infinitesimal magnetic
Reynolds number $Rm=\Omega r_i^2/\eta$, in which the induced magnetic field
is given by $Rm{\bf b}$, but $Rm$ no longer appears in the actual equations
to be solved.  In this $Rm\to0$ limit the Lorentz force
$(\nabla\times{\bf b})\times{\bf B}_0$ in (1), and the inductive term
$\nabla\times({\bf U\times B}_0)$ in (2) also become linear.  The only
nonlinearity is therefore the inertial term $Re\bf U\cdot\nabla\bf U$.

The boundary conditions associated with (1) are
$${\bf U}=\sin\theta{\bf\hat e}_\phi\quad{\rm at}\quad r=r_i,\qquad
  {\bf U}={\bf 0}\quad{\rm at}\quad r=r_o,\eqno(3)$$
where we fix $r_i=1$ and $r_o=2$.  The boundary conditions associated with
(2) are somewhat more complicated, and in particular depend on whether the
inner and outer regions are taken to be insulators or conductors.  There are
then four possible combinations, II, CI, IC, and CC, where the first letter
indicates the inner boundary and the second letter the outer.  We will present
only II and CC results here; the other two cases do not appear to yield
fundamentally new dynamics.  See \cite{Holl00a} for sample calculations of
all four, but only in the linear limit $Re\to0$.  For the actual mathematical
formulation of the II and CC boundary conditions, we refer to \cite{HS01},
where both are derived in detail.

These equations and associated boundary conditions are solved using the code
described by \cite{Holl00b}, in which the radial structure is expanded in
terms of Chebyshev polynomials, and the angular structure in terms of
spherical harmonics.  In this work we will consider only axisymmetric
solutions, so the code was reduced to a two-dimensional version, with no
variation in $\phi$.  In $r$ and $\theta$, resolutions as high as $150$ and
$700$ respectively were used.  Equation (1) is time-stepped by a second-order
Runge-Kutta method, modified to treat the diffusive term implicitly.
At each time-step of (1), (2) is inverted for $\bf b$.

The imposed field ${\bf B}_0$ is taken to be a linear combination
of dipolar and axial fields,
$${\bf B}_0=C({\bf B}_d+\epsilon{\bf B}_a),\qquad{\rm where}\qquad
{\bf B}_d=8r^{-3}\cos\theta\,\hat{\bf e}_r+4r^{-3}\sin\theta\,\hat{\bf e}_\theta,
\quad{\bf B}_a=\cos\theta\,\hat{\bf e}_r-\sin\theta\,\hat{\bf e}_\theta.\eqno(4)$$
Note how ${\bf B}_d$ and ${\bf B}_a$ have the same $\theta$ dependence; by
adjusting $\epsilon$ one can therefore impose $B_r=0$ on the particular
radial surface $r=-2/\epsilon^{1/3}$.  We will consider $\epsilon=-1$, $-8$,
and $-8/1.5^3$, yielding $B_r=0$ on the outer boundary, the inner boundary,
and at the midpoint $r=1.5$.  Finally, once $\epsilon$ has been specified, the
constant $C$ is adjusted such that the volume-averaged field strength is unity,
$\int_V|{\bf B}_0|^2dV/V=1$.  Its dimensional magnitude has after all already
been incorporated into the Hartmann number.  The field still has considerable
spatial variation though, so Hartmann numbers for different $\epsilon$ are not
necessarily directly comparable.

\section{Results}

\begin{figure}
\centering
\includegraphics[scale=0.5]{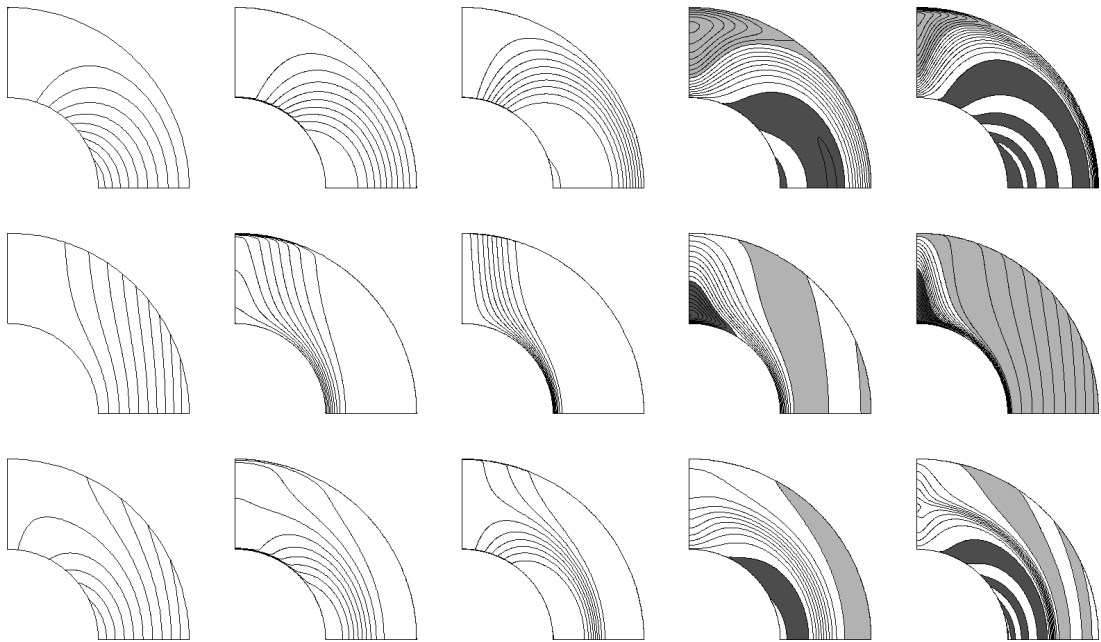}
\caption{From top to bottom, the three rows are at $\epsilon=-1$, $-8$,
and $-8/1.5^3$.  The first panel in each row shows the field lines of the
corresponding imposed fields ${\bf B}_0$; note in particular how $B_r=0$ at
$r=2$, 1, and 1.5, respectively.  The next four panels show contours of the
angular velocity $\omega$, with a contour interval 0.1, for the four cases
($Ha=10^2$, II), ($Ha=10^3$, II), ($Ha=10^2$, CC), ($Ha=10^3$, CC), and all
at $Re=0$.  For the CC results, dark-shaded regions are super-rotating
($\omega>1$), and light-shaded regions are counter-rotating ($\omega<0$).}
\end{figure}

We begin by reviewing the $Re\to0$ linear regime previously considered by
\cite{Holl00a,Holl01}.  Figure 1 shows both the imposed fields ${\bf B}_0$,
as well as the resulting angular velocities.  Focusing first on the simpler
II cases, we see that the angular velocity $\omega$ is essentially constant
on field lines of ${\bf B}_0$, as expected according to Ferraro's law of
isorotation \cite{Cowling}.  A Shercliff layer forms at the radius where
$B_r=0$.  Hartmann layers also form at boundaries where $B_r\neq0$.  The
Hartmann layers are much thinner than the Shercliff layers, scaling as
$Ha^{-1}$ versus $Ha^{-1/2}$.

Turning next to the CC cases, we see that there are now regions that are
super-rotating, with $\omega>1$, as well as counter-rotating, with $\omega
<0$.  With conducting boundaries, electric currents can recirculate through
them, leading to increased currents in the fluid, and hence increased
Lorentz forces, which accelerate the fluid into super- and counter-rotation.
The Shercliff layers are in the same locations as in the II cases, but are
considerably thinner than before (but still scale as $Ha^{-1/2}$).  The
Hartmann layers have been largely eliminated; the electro-magnetic coupling
across a conducting boundary is so strong that such boundary layers cannot
exist as before.

A few calculations were also done for CI and IC boundary conditions, with
the following results:  For $\epsilon=-1$, CI is similar to CC, and IC is
similar to II, whereas for $\epsilon=-8$, CI is similar to II, and IC is
similar to CC.  The classification is evidently determined by the boundary
where $B_r\neq0$, that is, by the boundary where a Hartmann layer forms for
I, but is suppressed for C.  In contrast, Shercliff layers can form at both
I and C boundaries (because they are not really boundary layers at all, but
rather internal shear layers that just happen to coincide with the
boundaries for the special cases $\epsilon=-1$ and $-8$).  It is not
surprising then that the boundary where $B_r\neq0$ is more important in
determining the similarities between different cases.

\begin{figure}
\centering
\includegraphics[scale=0.5]{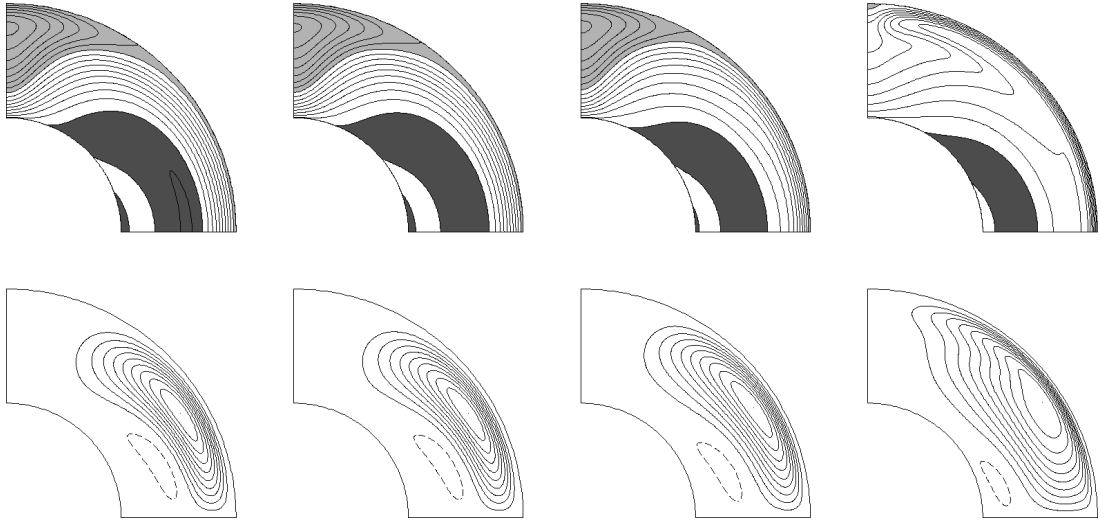}
\caption{The $\epsilon=-1$, CC solutions, for $Ha=10^2$, and $Re=30$, 100,
300, and 1000, from left to right.  The top row shows contours of the
angular velocity, as in Figure 1.  The bottom row shows streamlines of the
meridional circulation $\psi$, with solid lines denoting counter-clockwise
flow, and dashed lines clockwise.  From left to right, the maximum values
of $\psi$ are $7.26\times10^{-3}$, $2.33\times10^{-2}$, $5.14\times10^{-2}$, and $5.13\times10^{-2}$.}
\end{figure}

\begin{figure}
\centering
\includegraphics[scale=0.5]{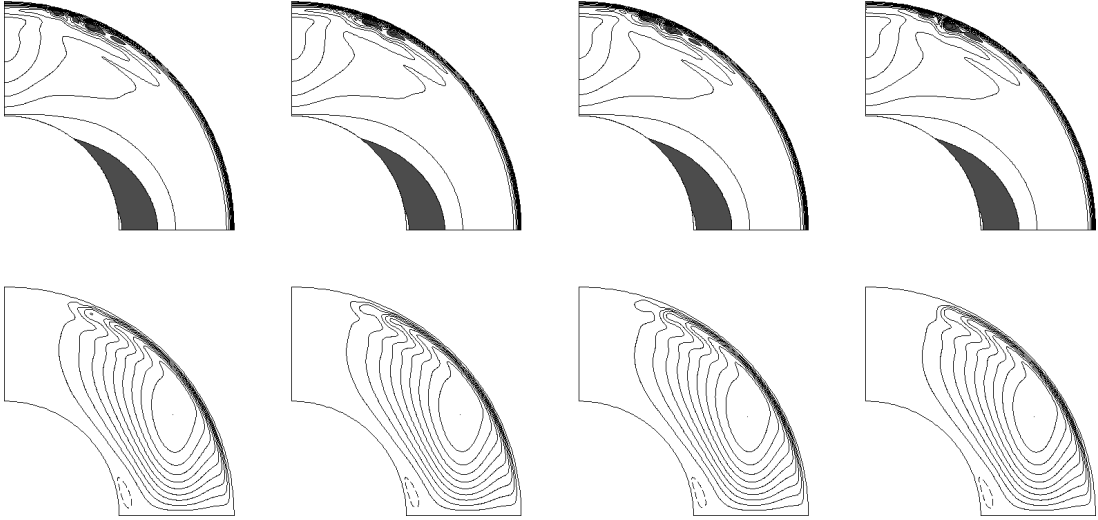}
\caption{The $\epsilon=-1$, CC solution, for $Ha=10^2$ and $Re=3000$.
From left to right, four snapshots of the time-dependent solution, uniformly
spaced throughout the period $\tau=27$.  As in Figure 2, the top row shows
contours of $\omega$, the bottom row $\psi$, with $\psi_{max}=3.56\times10^{-2}$.}
\end{figure}

In the remainder of this paper we wish to consider the influence of
increasingly large $Re$ on these solutions in Figure 1.  Figure 2 shows the
$\epsilon=-1$, CC results, for $Ha=10^2$ and $Re$ from 30 to 1000.  One new
feature that immediately emerges as soon as $Re>0$ is a meridional circulation.
This acts back on the angular velocity in a number of ways.  First, because it
is outward near the equator, it tends to compress the Shercliff layer against
the outer boundary, so that by $Re=1000$ it is much thinner than it
was for $Re=0$.  Second, we note that by $Re=1000$ the counter-rotation has
been almost completely eliminated, and the super-rotation has also been
considerably reduced.

In Figure 3, $Re$ is further increased to 3000, which results in a
time-dependent solution.  The time-dependence is almost exclusively confined
to the Shercliff layer though, consisting of a series of ripples within this
layer.  It is quite similar to the time-dependent solutions previously
obtained in a purely dipolar field \cite{HCF07}, even though the boundary
layer in that case was a true Hartmann layer rather than the Shercliff layer
here.  For $Re>0$ the distinction between Hartmann and Shercliff layers is
apparently less critical than it is for $Re=0$.

For $Ha=10^2$, the onset of this time-dependence is at $Re_c\approx2200$,
and appears to be a subcritical Hopf bifurcation.  That is, if one gradually
increases $Re$ until the onset of instabilities, and then decreases it again,
there is a certain degree of hysteresis before one switches back to the
steady solutions.  These calculations are unfortunately too time-consuming
to allow a precise determination of the initial $Re_c$, or the extent of
the hysteresis.

If we next increase the Hartmann number to $Ha=10^{2.5}$, the onset of
time-dependence is at $Re_c\approx6000$ (again with hysteresis), perhaps
suggesting an $Re_c\sim Ha$ scaling.  Precisely determining how $Re_c$
scales with $Ha$ would obviously be even more time-consuming though than
simply determining it at any particular value of $Ha$.  Further increasing
$Ha$ to $10^3$, the solutions are (not surprisingly) stable at least up to
$Re=10000$.  Figure 4 shows results for $Ha=10^3$ and $Re$ from 300 to
10000.  One very interesting result emerges from comparing the two $Re=1000$
solutions in Figures 2 and 4: the thickness of the Shercliff layer now
appears to be controlled far more by $Re$ than by $Ha$.  There are
unfortunately no analytic predictions to compare with in this nonlinear
regime.

\begin{figure}
\centering
\includegraphics[scale=0.5]{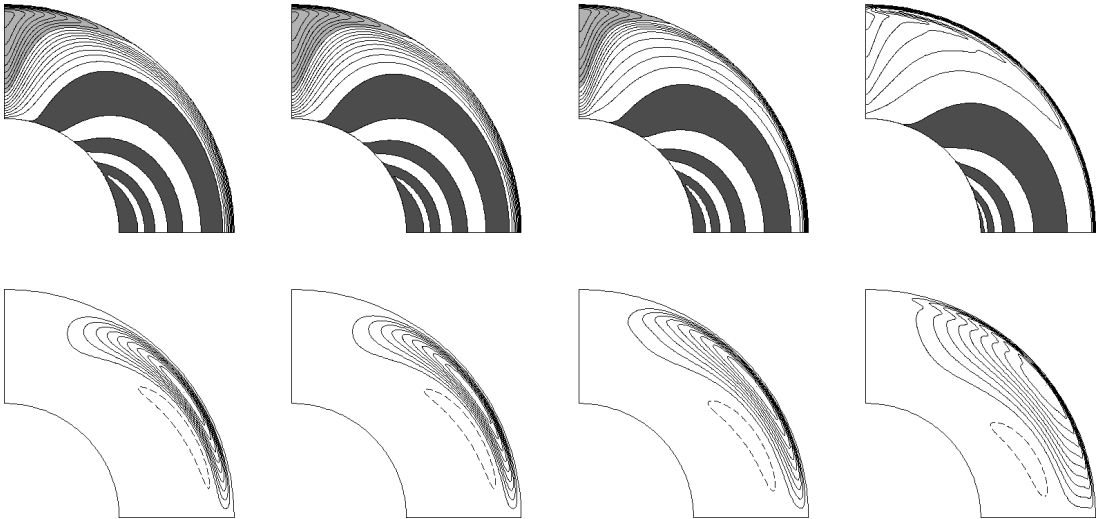}
\caption{The $\epsilon=-1$, CC solutions, for $Ha=10^3$, and $Re=300$,
1000, 3000, 10000, from left to right.  As in Figure 2, the top row shows
contours of $\omega$, the bottom row $\psi$, with $\psi_{max}=4.17\times10^{-3}$, $1.27\times10^{-2}$, $2.18\times10^{-2}$, and $1.72\times10^{-2}$.}
\end{figure}

\begin{figure}
\centering
\includegraphics[scale=0.5]{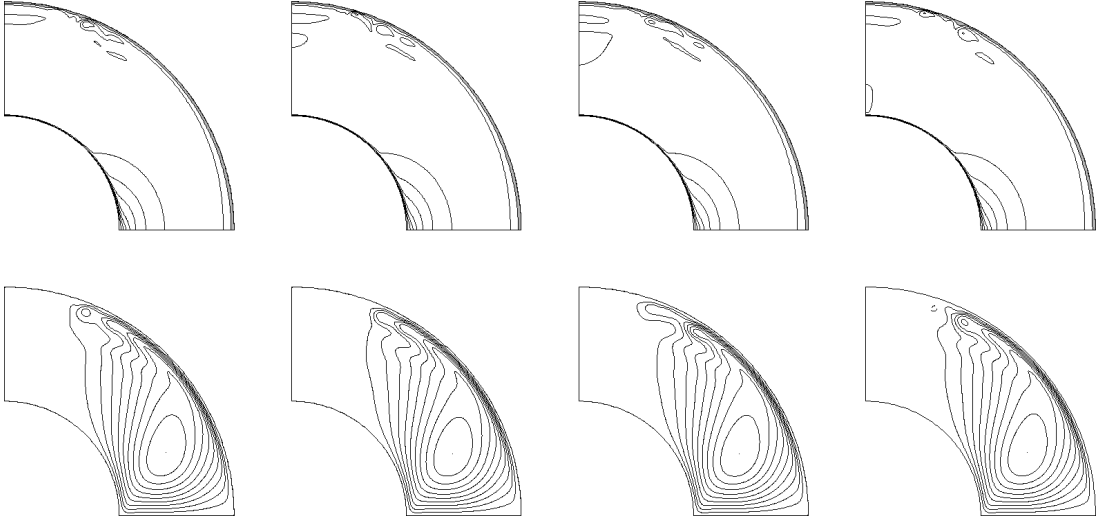}
\caption{The $\epsilon=-1$, II solution, for $Ha=10^2$ and $Re=10000$.
From left to right, four snapshots of the time-dependent solution, uniformly
spaced throughout the period $\tau=20$.  As in Figure 2, the top row shows
contours of $\omega$, the bottom row $\psi$, with $\psi_{max}=1.31\times10^{-2}$.}
\end{figure}

Figure 5 shows the result of switching from CC to II boundary conditions.
We still obtain a time-depen\-dence very similar to that in Figure 3, but the
onset does not occur until $Re_c\approx6500$, again with hysteresis.  II
boundaries are therefore considerably more stable than CC.  The reason for
this is straightforward: For CC, the Hartmann layer on the inner boundary is
completely suppressed, so all of the adjustment in angular velocity occurs
across the Shercliff layer at the outer boundary.  In contrast, for II, much
of the adjustment occurs across the Hartmann layer, so the Shercliff layer
is considerably weaker than for CC.  A weaker Shercliff layer then requires
a much larger Reynolds number before it will become unstable.

\begin{figure}
\centering
\includegraphics[scale=0.5]{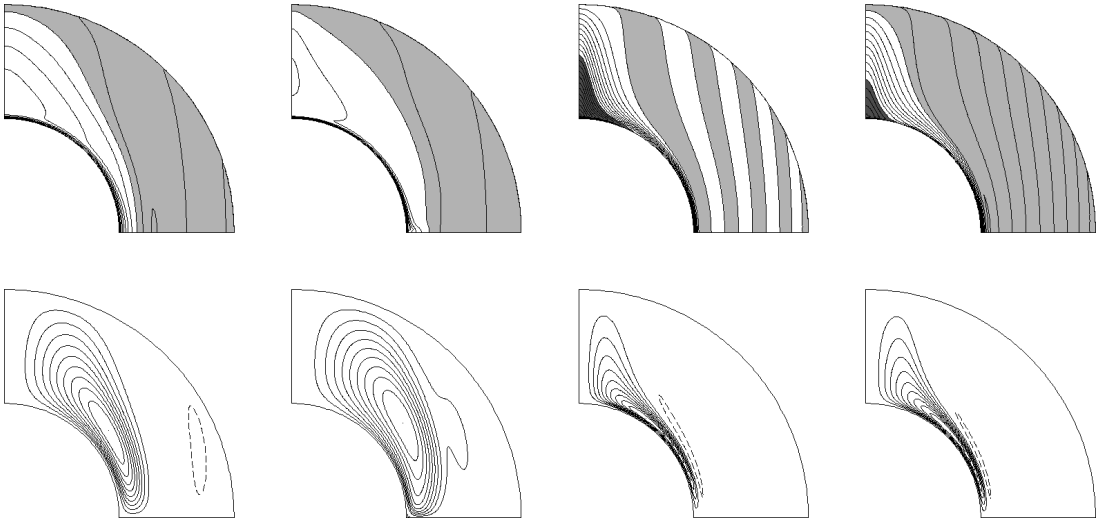}
\caption{The $\epsilon=-8$, CC solutions, from left to right
($Ha=10^2$, $Re=3000$), ($Ha=10^2$, $Re=10000$), ($Ha=10^3$, $Re=3000$),
($Ha=10^3$, $Re=10000$).  As in Figure 2, the top row shows contours of
$\omega$, the bottom row $\psi$, with $\psi_{max}=1.40\times10^{-2}$, $1.24\times10^{-2}$, $1.26\times10^{-3}$, and $2.93\times10^{-3}$.}
\end{figure}

We next switch to $\epsilon=-8$.  Figure 6 shows these solutions, at $Ha=10^2$
and $10^3$, and $Re=3000$ and 10000 for each.  Increasing $Ha$ again increases
the super-/counter-rotation, whereas increasing $Re$ suppresses them.  In this
case the inner Shercliff layer is very stable though, with no evidence of any
transition to time-dependence even up to $Re=30000$.  Partly this may be due
to the spatial variation of ${\bf B}_0$; $B_\theta$ at the inner boundary for
$\epsilon=-8$ is $8\sqrt{5/68}=2.2$ times stronger than $B_\theta$ at the
outer boundary for $\epsilon=-1$.  It seems very unlikely though that a mere
2.2 times increase in field strength (for a given Hartmann number) could by
itself cause such an enormous increase in critical Reynolds number, from
$\sim2200$ to over 30000.  We conclude therefore that an inner Shercliff
layer is intrinsically more stable than an outer Shercliff layer.  The
instabilities of the outer Shercliff layer are perhaps similar to G\"ortler
vortices, which also occur only if a boundary is curved one way, but not the
other.

Finally, Figures 7 and 8 show results for $\epsilon=-8/1.5^3$.  Increasing
$Re$ in this case appears to destroy the Shercliff layer completely.  For
sufficiently large $Re$ these solutions again become time-dependent, but
the instabilities are no longer confined to boundary layers.  One further new
feature that emerges for sufficiently large $Re$ is a super-rotating region
near the pole.  This is quite different from the previous super-rotations
though, which were all driven by the Lorentz force.  In contrast, this
super-rotation is induced by the inertial term: as the meridional circulation
sweeps fluid inward toward the pole, conservation of angular momentum causes
its angular velocity to increase, eventually exceeding one.

\begin{figure}
\centering
\includegraphics[scale=0.5]{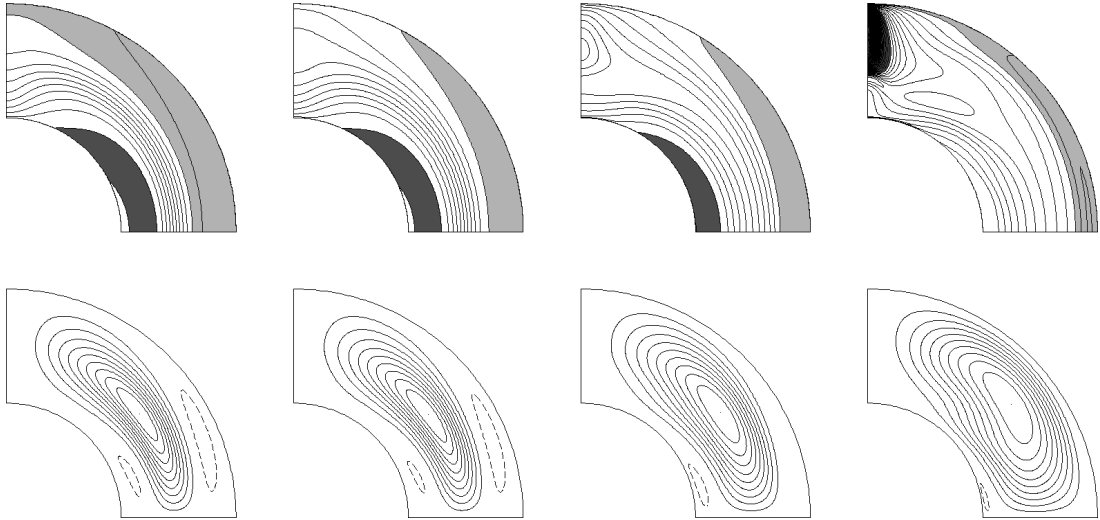}
\caption{The $\epsilon=-8/1.5^3$, CC solutions, for $Ha=10^2$, and $Re=100$,
300, 1000, 3000, from left to right.  As in Figure 2, the top row shows
contours of $\omega$, the bottom row $\psi$, with $\psi_{max}=7.71\times10^{-3}$, $2.15\times10^{-2}$, $4.55\times10^{-2}$, and $6.64\times10^{-2}$.}
\end{figure}

\begin{figure}
\centering
\includegraphics[scale=0.5]{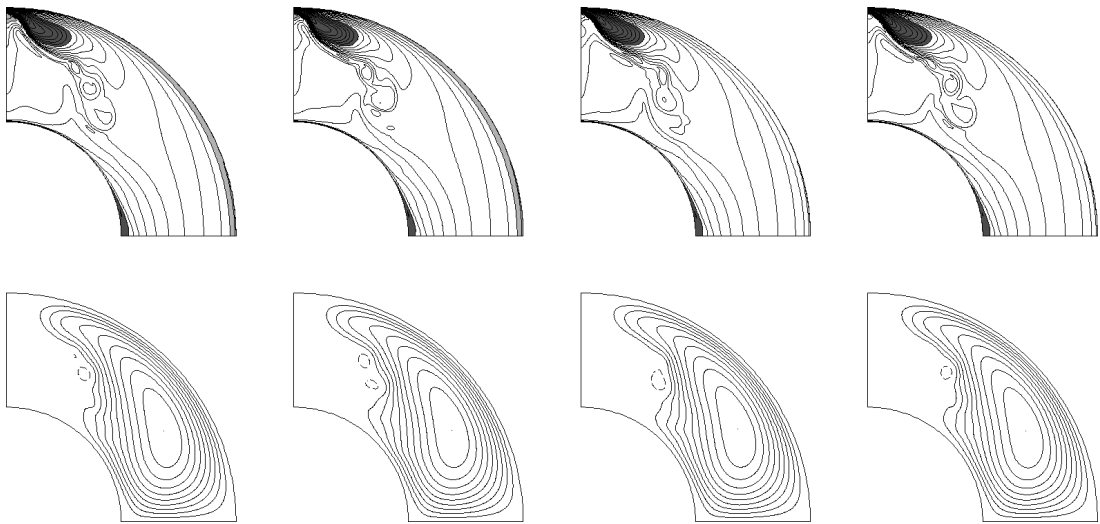}
\caption{The $\epsilon=-8/1.5^3$, CC solution, for $Ha=10^2$ and $Re=10000$.
From left to right, four snapshots of the time-dependent solution, uniformly
spaced throughout the period $\tau=10$.  As in Figure 2, the top row shows
contours of $\omega$, the bottom row $\psi$, with $\psi_{max}=7.44\times10^{-2}$.}
\end{figure}

\section{Conclusion}

In this work we have considered magnetic spherical Couette flow in three
particular field configurations ${\bf B}_0$, specifically chosen to place
Shercliff layers at either the outer boundary, or the inner, or exactly in
between.  In each case we then increased the Reynolds number, and studied
the resulting balance and competition between magnetic and inertial effects.
We found that increasing the Hartmann number stabilizes the solutions,
whereas increasing the Reynolds number tends to de-stabilize them, except
for $\epsilon=-8$, where we never obtained any instabilities.  It would be
of interest to further increase $Re$ in this case; some type of instabilities
must surely set in eventually, but not necessarily boundary layer
instabilities of the type obtained for $\epsilon=-1$.  Future work should
also explore the possibility of non-axisymmetric instabilities, as in the
purely axial field case \cite{HS01,Xing,Holl09}.

Finally, returning to the magnetic spherical Couette flow experiments that
have previously been done \cite{Dan,DTS1,DTS2}, it would be of considerable
interest to extend one or the other of these to impose not just purely
axial or purely dipolar fields, but some of the configurations presented
here.  We hope our results may stimulate future experimental work in this
direction.

\begin{acknowledgements}
We thank Prof. Andrew Jackson for providing the computational facilities.
\end{acknowledgements}

\end{document}